# Wafer-scale single-crystal hexagonal boron nitride monolayers on Cu (111)

Tse-An Chen[1+], Chih-Piao Chuu[1+], Chien-Chih Tseng[2], Chao-Kai Wen[2], H.-S. Philip Wong[1], Shuangyuan Pan[3], Rongtan Li [4,5], Tsu-Ang Chao[1], Wei-Chen Chueh[2], Yanfeng Zhang[3], Qiang Fu[4], Boris I. Yakobson[6]*, Wen-Hao Chang[2,7]*, Lain-Jong Li[1]*

[1]Corporate Research, Taiwan Semiconductor Manufacturing Company (TSMC), 168 Park Ave. 2, Hsinchu Science Park, Hsinchu 30075, Taiwan.

[2]Department of Electrophysics, National Chiao Tung University, Hsinchu 30010, Taiwan.

[3]Department of Materials Science and Engineering, College of Engineering, Peking University, Beijing 100871, China.

[4]State Key Laboratory of Catalysis, Dalian Institute of Chemical Physics, the Chinese Academy of Sciences, Dalian 116023, China.

[5]University of Chinese Academy of Sciences, Beijing 100049, China.

[6]Department of Materials Science and Nanoengineering, Department of Chemistry, and the Smalley Institute for Nanoscale Science and Technology, Rice University, Houston, Texas 77005, United States.

[7]Center for Emergent Functional Matter Science (CEFMS), National Chiao Tung University, Hsinchu 30010, Taiwan.

[+] These authors contribute equally.

*To whom correspondence should be addressed: ljliv@tsmc.com, whchang@mail.nctu.edu.tw or biy@rice.edu.

**Ultrathin two-dimensional (2D) semiconducting layered materials offer a great potential to extend the Moore's Law (*1*). One key challenge for 2D semiconductors is to avoid the formation of charge scattering and trap sites from adjacent dielectrics. The insulating van der Waals layer, hexagonal boron nitride (*h*BN), is an excellent interface dielectric to 2D semiconductors, efficiently reducing charge scatterings (*2, 3*). Recent studies have shown the growth of single-crystal *h*BN films on molten Au surfaces (*4*) or bulk Cu foils (*5*). However, using molten Au is not favored in industry due to high cost, cross-contamination, and potential issues of process control and scalability. Cu foils may be suitable for roll-to-**

roll processes, but unlikely to be compatible with advanced microelectronic fabrication on Si wafers. Thus, only a reliable approach to grow single-crystal *h*BN on wafers can help realize the broad adoption of 2D layered materials in industry. Previous efforts on growing *h*BN triangular monolayers on Cu (111) metals have failed to achieve mono-orientation, resulting in unwanted grain boundaries when they merge as films (*6,7*). Growing single-crystal *h*BN on such a high-symmetry surface planes (*5,8*) is commonly believed to be impossible even in theory. In stark contrast, we have successfully realized the epitaxial growth of single-crystal *h*BN monolayers on a Cu (111) thin film across a 2-inch *c*-plane sapphire wafer. This surprising result is corroborated by our first-principles calculations, suggesting that the epitaxy to the underlying Cu lattice is enhanced by the lateral docking to Cu (111) steps, to ensure the mono-orientation of *h*BN monolayers. The obtained single-crystal *h*BN, incorporated as an interface layer between $MoS_2$ and $HfO_2$ in a bottom-gate configuration, has enhanced the electrical performance of transistors based on monolayer $MoS_2$. This reliable approach of producing wafer-scale single-crystal *h*BN truly paves the way for developing futuristic 2D electronics.

First, a single-crystal Cu (111) thin film on a wafer is needed. Single-crystal Cu in thick foils can be achieved through recrystallization induced by implanted seeds (*5,9*). However, for the formation of Cu (111) thin film on a wafer, the crystallinity strongly relies on the underlying substrate lattices. Here we used a *c*-plane sapphire as the substrate, on which a 500-nm-thick polycrystalline Cu film was sputtered followed by extensive thermal annealing to achieve single-crystal Cu (111) films (*10*). One challenge is that Cu (111) tends to form twin grains separated by twin grain boundaries, through kinetic growth processes. **Fig. 1a** illustrates the atomic arrangements for the typical twinned Cu (111) structure. We find that the post-annealing at a high temperature (1,040 - 1,070 °C) in the presence of hydrogen is the key to removing the twin grains, consistent with recent reports (*10,11*). **Figures 1b and 1c** show the optical micrographs (OMs) and electron backscatter diffraction (EBSD) patterns for the Cu (111) thin films after annealing at 1,000 °C and 1,050 °C, respectively. The EBSD results (see also Extended Data Fig. 1a-b) conclude the coexistence of twinned Cu (111) polycrystals with 0° and 60° in-plane misorientation for the Cu thin films annealed at 1,000 ºC. The in-plane misorientation was removed after annealing at 1,050 ºC, which evidences the formation of single-crystal Cu (111).

The X-ray diffraction results (Extended Data Fig. 1c-f) also consistently prove the success in obtaining single-crystal Cu (111) thin films. Note that the Cu (111) is preferentially formed with a thinner Cu film, but a sufficiently thick Cu film is necessary to prevent Cu evaporation during the subsequent *h*BN growth. Thus, there exists an optimal Cu thickness at around 500 nm for our single-crystal *h*BN growth.

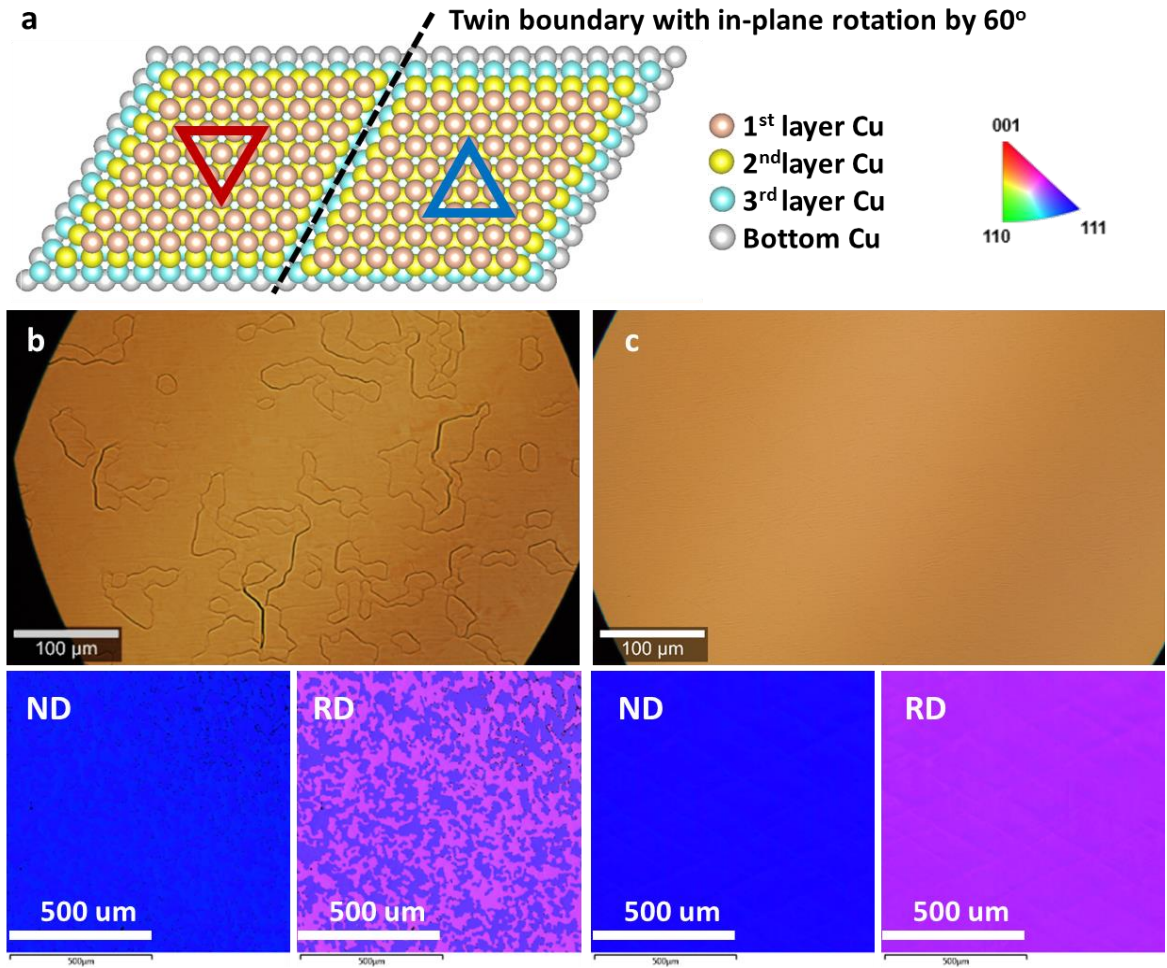

**Fig. 1 | Cu (111) lattice orientation on c-sapphire substrates. a**, Schematics of twinned Cu (111) (top view) with an in-plane rotation by 60°, where the blue and red triangles enclose the areas with the same Cu stacking configurations. **b**, Optical microscope image and EBSD inverse pole figure (IPF) mappings (1 × 1 mm$^2$) of the Cu thin film annealed at 1,000 °C for 1 hr, where the normal direction (ND) and rolling direction (RD) indicate the presence of twinned Cu (111) polycrystals with in-plane misorientation by 60°. **c**, The film annealed at 1,050 °C for 1 hr unifies the in-plane orientation, evidencing the formation of single-crystal Cu (111) without any twinned grains.

Achieving the growth of mono-oriented *h*BN triangular flakes is a critical step towards obtaining wafer-scale single-crystal *h*BN. Owing to the six-fold symmetry of Cu (111), the van der Waals (vdW) registry of *h*BN to Cu (111) leads to two sets of energy-minimal configurations (orientation differs by 60° or 180°, an inversion) with almost degenerate binding energies. Thus it is commonly believed that confining *h*BN flakes to mono-orientation on such high-symmetry surface is impossible (*5,8*). Our experiments point out that the energy degeneracy can be lifted in the presence of spontaneously present top-layer Cu step edges. The growth of *h*BN monolayer is performed by flowing ammonia borane precursors to the one-inch single-crystal Cu (111) thin film/sapphire in a hot wall CVD furnace. The OM image for the monolayer *h*BN triangular flakes grown on a Cu (111) thin film with twin grains (**Figure 2a**) demonstrates that *h*BN flakes orient to the same direction on one twin and to opposite direction (or 60° in-plane rotation along the z-axis) on the counterpart twin (Extended Data Fig. 2). **Figure 2b** displays the OM images for the *h*BN flakes grown on a single-crystal Cu (111) thin film without twin grains, where almost all triangles are unidirectionally aligned (see also Extended Data Fig. 3 for statistical analysis for the orientation distribution of *h*BN triangular flakes). The observation of mono-orientation on an individual single-crystal Cu (111) grain clearly indicates the existence of an energy-minimized *h*BN-Cu (111) configuration. Therefore, eliminating the twin grains in Cu (111) shall ensure the growth of single-crystal *h*BN on it. To verify the single crystallinity, the *h*BN monolayer merged from mono-oriented triangles has been characterized by micro-spot low-energy electron diffraction (*μ*-LEED) with a probe size around 3 μm at 80 sites across the one-inch wafer. **Figure 2c** displays the *μ*-LEED patterns from 9 randomly selected sites. All graphs demonstrate a unidirectionally aligned *h*BN monolayer with the Cu (111) surfaces, indicating their single crystallinity strictly following the Cu (111) lattices. The atomically resolved scanning tunneling microscopy (STM) image for *h*BN on Cu (111) in **Fig. 2d** demonstrates perfect *h*BN lattice with measured lattice constant $2.50 \pm 0.1$ Å, consistent with the theoretical value 2.5 Å. We have probed more than 20 locations and all STM images show the same *h*BN lattice orientation (Extended Data Fig. 4). We did not observe any grain boundaries formed by adjacent mis-oriented *h*BN domains, suggesting the single-crystalline nature of *h*BN. Note that in some areas, moiré patterns arise from the lattice mismatch and/or relatively small rotation (within 1.5°) between *h*BN and underlying Cu (111) substrate (Extended Data Fig. 5a-f). The magnified atomic resolution image at the moiré boundary areas reveals that the *h*BN presents perfect lattice coherence at the patching boundary (Extended Data Fig. 5g-j), indicating that the formation of moiré patterns does not affect the overall hBN orientation. We believe that the hBN complete the single-crystal growth at high temperatures and the strain associated with the sample cooling after growth results in the formation of local moiré pattern. Other characterizations including X-ray photoelectron spectroscopy (XPS) and Raman spectroscopy prove the B-N chemical bonding structures (**Figs. 2e-2f**). The transmission electron microscope (TEM) and atomic force microscope (AFM) images consistently show that the as-grown hBN is indeed a monolayer (**Figs. 2g-2h**).

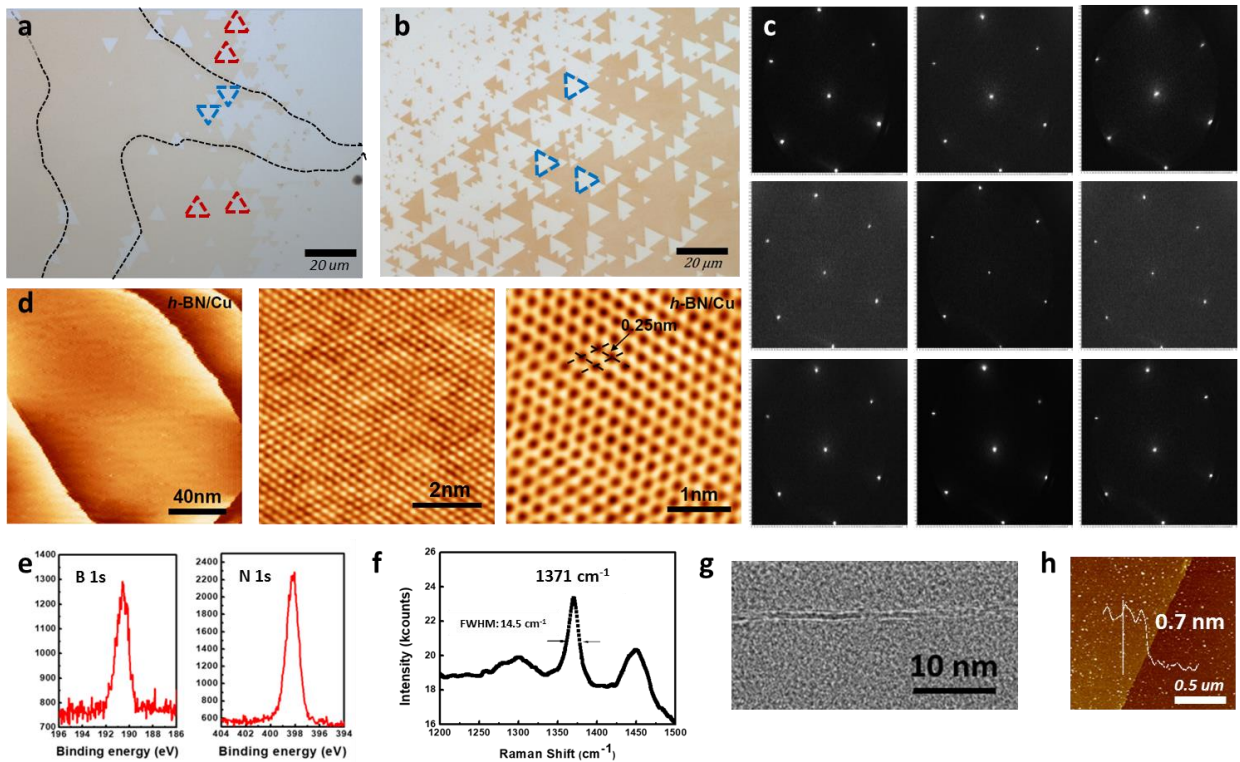

**Fig. 2 | Growth and atomic structures of single-crystal hBN on Cu (111) films. a**, Optical microscope image of the hBN grown on different Cu (111) grains (black dashed lines indicate the twin grain boundaries), where oppositely-oriented flakes were marked by red and blue dashed triangles. **b**, Mono-oriented hBN flakes on single-crystal Cu (111) films. **c**, $\mu$-LEED patterns of hBN monolayers at 9 different areas randomly selected from the 1.5 × 1.5 cm$^2$ sample surface. All $\mu$-LEED patterns show that the hBN monolayers have the same orientation as the Cu (111) surfaces. **d**, Left to right: Large-scale STM image ($V_{Tip}$ = −1.008 V, $I_{Tip}$ = 3.90 nA, T = 300 K) and atomic-scale STM image ($V_{Tip}$ = −0.003 V, $I_{Tip}$ = 54.508 nA, T = 300 K) of hBN/Cu (111) with measured lattice constant 2.50 ± 0.1Å. Note the height of each step is measured as ~2Å. **e**, XPS spectra measured from an as-grown monolayer hBN film on Cu (111)/sapphire. The binding energies of B 1s and N 1s at 190.4 eV and 398.0 eV confirm the formation of hBN. **f**, Raman spectrum of a transferred hBN film, where the $E_{2g}$ mode at 1371 cm$^{-1}$ with a full width at

half maximum (FWHM) 14.5 cm$^{-1}$ confirms that the *h*BN is a monolayer. **g**, Cross-sectional TEM image of a monolayer *h*BN transferred on SiO$_2$/Si, where the thickness of *h*BN is around 0.5 nm. **h**, AFM image of a single-crystal *h*BN film transferred onto SiO$_2$/Si substrate.

We recognize that once the Cu (111) thin films are prepared and formed untwined at 1,050 °C as described above, the subsequent mono-orientated growth of *h*BN flakes can be realized at various growth temperatures ranging from 995 to 1,070 °C, as displayed in Extended Data Fig. 6. However, a lower growth temperature (995 to 1,010 °C) usually leads to lower-quality *h*BN flakes, which is easily oxidized by subsequent oxidation test at 150 °C in air. Therefore, we used a higher growth temperature (typically 1,050 °C) to ensure high-quality single-crystal *h*BN growth.

To explain the preferred orientation of *h*BN on Cu (111), we consider a small and rigid B$_6$N$_7$ molecule [i.e., an energetically favorable N-terminated three-ring structure (*12*)] as the probe-seed. We first examine the effect of plane-to-plane epitaxy, using density functional theory (DFT) to calculate the binding energies of 6 typical atomic stacking configurations as shown in **Fig. 3a**, where N$_I$B$_{III}$, N$_{III}$B$_{II}$ and N$_{II}$B$_I$ are defined as the 0° orientation, whereas N$_I$B$_{II}$, N$_{II}$B$_{III}$ and N$_{III}$B$_I$ are the 60° (that is inverted) orientation. The notation N$_i$B$_j$ represents the stacking of N atoms in registry with (above) the Cu atoms in the i$^{th}$ layer, while B atoms register with the Cu atoms in the j$^{th}$ layer. The calculations show that the stacking with N atoms on top of the first-layer Cu atoms [N$_I$B$_{III}$ (0°) and N$_I$B$_{II}$ (60°)] exhibits the lowest energy, while B atoms on top of the first-layer Cu atoms [N$_{II}$B$_I$ (0°) and N$_{III}$B$_I$ (60°)] are energetically unfavorable. The preferential registrations reflect the electron affinity of the B and N atoms, which leads to attractive (repulsive) Coulomb interactions between N (B) atoms and the first-layer Cu atoms, and hence affects the structural stability. We find that the lowest-energy structures for 0° (N$_I$B$_{III}$) and 60° (N$_I$B$_{II}$) orientations exhibit only an energy difference ~0.05 eV, significantly smaller than the thermal energy k$_B$T at the growth temperature (~0.1eV), indicating that the plane-to-plane registry is insufficient to achieve mono-oriented growth, in agreement with reported simulations (*13*).

Actually the Cu (111) surface is not perfectly flat and many terrace meandering steps exist as revealed in STM images (**Fig. 2d** and Extended Data Fig. 7a-d). A recent theory work pointed out that one must consider the guidance from these step edges on the *h*BN growth (*14*).

Wang *et al.* argued that docking at the vicinal step edges on Cu (110) surface governs the single-crystal *h*BN growth, relying on the assumption that the Cu terrace steps only trend up or down all the way across the whole vicinal surface of the Cu foil. Our STM results clearly reveal that the terrace steps of Cu (111) surface trend both up and down across the wafer, and the edge-docking can seemingly yield *h*BN in both directions, unless the binding energies differ enough to favor one direction over another. To capture this into our model, an extra layer of Cu atoms (red in **Fig. 3**) is added on top of the 1st layer, forming two opposite step edges (A- and B-step edges shown in **Fig. 3a**, and Extended Data Fig. 7f). This restricts the $B_6N_7$ seed to the 0° (60°) orientation when docking to the A- (B-) steps (**Fig. 3a**). In **Figs. 3a** and **3b** the distance between the Cu step edge and the $B_6N_7$ zigzag edge for each configuration has been determined by energy minimization as detailed in Extended Data Fig. 8. The resulting binding energy for each configuration has a subtle yet very important change with the presence of Cu step edges, as displayed in **Fig. 3b**: the two configurations $N_IB_{II}$ (60°) and $N_IB_{III}$ (0°), which were nearly degenerate when considering only plane-to-plane epitaxy, are now separated by 0.23 eV, this value raising in proportion with the docking length, rapidly amplifying the Boltzmann selectivity factor (for just 5-6 hexagons contact to above $10^3$, Extended Data Fig. 9). Such an energy difference apparently ensures the mono-orientation growth. Our STM results in Extended Data Fig.7 show that all meandering steps are rather curved and locally rugged, so they all consist of segments of A and B types. BN seeds should kinetically nucleate at the A to B corner while docking to stronger binding sites, B-types, with proper orientation (Extended Data Fig. 7e). The simulations together with experimental results indicate the Cu (111) surfaces with the step edge formation are the key to achieving single-crystal *h*BN growth.

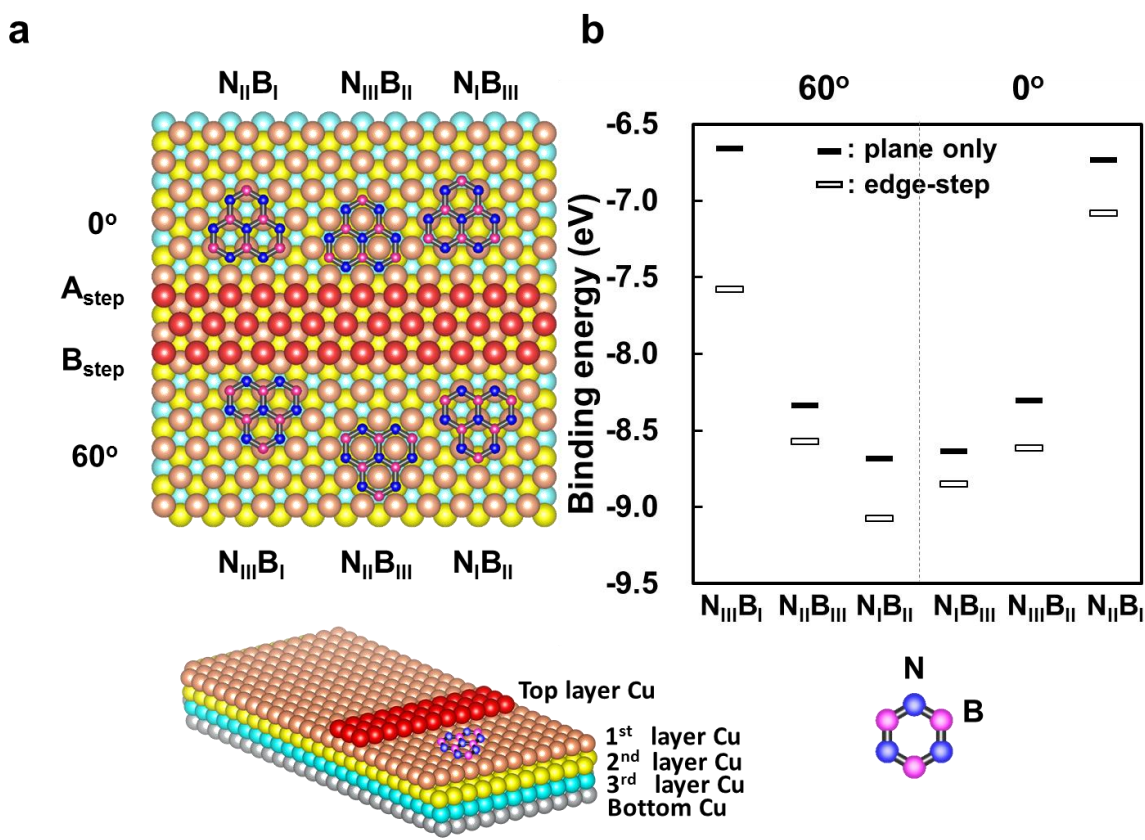

**Fig. 3 | DFT calculations for epitaxy at nucleation, with and without considering step-edge docking. a**, The lowest energy atomic arrangements for six $B_6N_7$-Cu (111) configurations considering the edge docking to two typical step edge termination (A and B) of top-layer Cu (111). **b**, Calculated binding energies for the six $B_6N_7$-Cu (111) configurations with and without including the edge-to-step epitaxy effect.

With the success in growth of single-crystal $h$BN on a one-inch Cu (111) thin film, we further scale the growth to a 2-inch wafer as depicted in **Fig. 4a**. Since the interaction between fully grown $h$BN layer and Cu (111) is limited to weak vdW forces, the detachment of wafer-scale $h$BN can be achieved by polymer-assisted transfer with the help of electrochemical processes (*15,16*) as schematically illustrated in **Fig. 4b** and Extended Data Fig. 10. **Figure 4c** shows the photo of a 2-inch $h$BN monolayer transferred onto a 4-inch $SiO_2$/Si wafer. We show that the growth of single-crystal wafer-scale $h$BN on Cu (111) thin films on wafers is scalable and much more cost-effective than using thick Cu foils or other novel metals; and thus is a preferred approach for the microelectronics industry. The availability of wafer-scale single-crystal $h$BN shall stimulate and enable further research and development of futuristic 2D electronics. We constructed monolayer MoS₂ FETs with and without single-crystal and polycrystalline *h*BN as an interface dielectric in a bottom-gate configuration as illustrated in the Extended Data Fig.11. The mobility enhancement of MoS₂ and hysteresis suppression are significant in the device integrated with a single-crystal *h*BN monolayer, indicating its promise for 2D-based transistors.

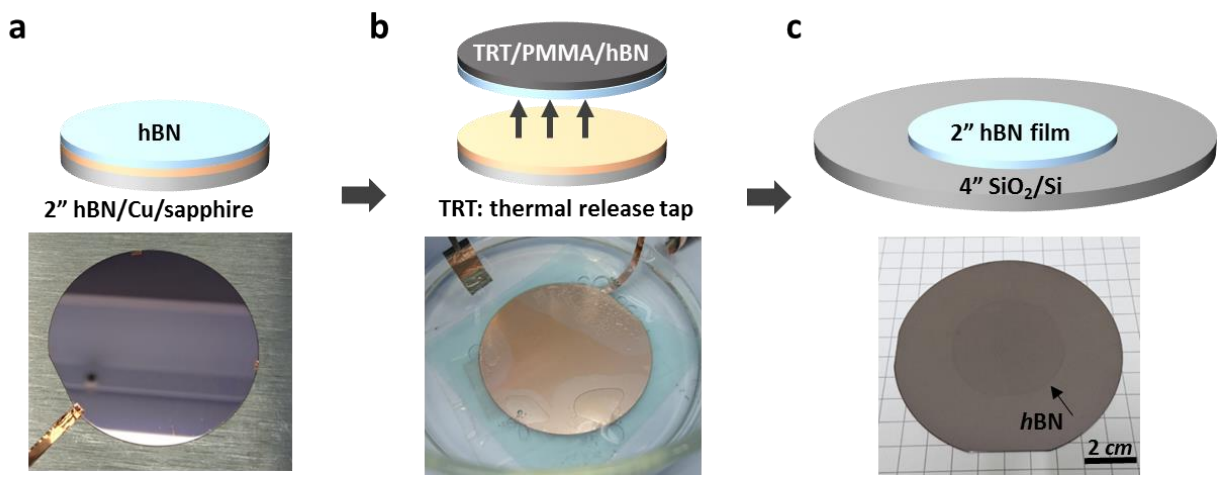

**Fig. 4 | Schematic and photos illustrating the wafer-scale *h*BN transfer processes. a**, As-grown 2-inch *h*BN film on a Cu (111)/sapphire wafer. **b**, Electrochemical delamination in an aqueous solution of NaOH (1M) as the electrolyte, using the Cu layer in the TRT/PMMA/*h*BN/Cu (111)/sapphire stack as the cathode and a platinum (Pt) foil as the anode with an applied DC voltage of 4 V. During this process, the TRT/PMMA/*h*BN stacked film was detached from the Cu (111)/sapphire through the generations of hydrogen bubbles at the *h*BN/Cu interfaces. **c**, Two-inch monolayer *h*BN film transferred onto a four-inch SiO₂/Si wafer.

**References and Notes:**

1. Li, M.-Y., Su, S.-K., Wong, H.-S. Philip & Li, L.-J. How 2D semiconductors could extend Moore's law. *Nature* **567**, 169-170 (2019).

2. Novoselov, K. S., Mishchenko, A., Carvalho, A. & Neto, A. H. C. 2D materials and van der Waals heterostructures. *Science* **353**, 9439 (2016).

3. Dean, C. R. et al. Boron nitride substrates for high-quality graphene electronics. *Nat. Nanotechnol.* **5**, 722–726 (2010).

4. Lee, J. S. et al. Wafer-Scale Single-Crystal Hexagonal Boron Nitride Film via Self-Collimated Grain Formation. *Science* **362**, 817–821 (2018).

5. Wang, L. et al. Epitaxial growth of a 100-square-centimetre single-crystal hexagonal boron nitride monolayer on copper. *Nature* **570**, 91–95 (2019)

6. Uchida, Y., Iwaizako, T., Mizuno, S., Tsuji, M. & Ago, H. Epitaxial Chemical Vapour Deposition Growth of Monolayer Hexagonal Boron Nitride on a Cu (111)/Sapphire Substrate. *Phys. Chem. Chem. Phys.* **19**, 8230−8235 (2017).

7. Song, X. et al. Chemical vapor deposition growth of large-scale hexagonal boron nitride with controllable orientation. *Nano Res.* **8**, 3164-3176 (2015).

8. Li, J. et al. Growth of Polar Hexagonal Boron Nitride Monolayer on Nonpolar Copper with Unique Orientation. *Small* **12**, 3645 (2016).

9. Jin, S. et al. Colossal grain growth yields single-crystal metal foils by contact-free annealing. *Science* **362**, 1021–1025 (2018).

10. Verguts, K. et al. Epitaxial Al$_2$O$_3$ (0001)/Cu (111) Template Development for CVD Graphene Growth. *J. Phys. Chem. C* **120**, 297-304 (2015).

11. Deng, B. et al. Wrinkle-free single-crystal graphene wafer grown on strain-engineered substrates. *ACS Nano* **11**, 12337-12345 (2017).

12. Liu, Y., Bhowmick, S. & Yakobson, B. I. BN White Graphene with "Colorful" Edges: The Energies and Morphology. *Nano Lett.* **11**, 3113–3116 (2011).

13. Zhao, R., Zhao, X., Liu, Z., Ding, F. & Liu, Z. Controlling the orientations of *h*BN during growth on transition metals by chemical vapor deposition. *Nanoscale* **9**, 3561–3567 (2017).

14. Bets, K. V., Gupta, N. & Yakobson, B. I. How the complementarity at vicinal steps enables growth of 2D monocrystals. *Nano Lett.* **19**, 2027–2031 (2019).

15. Gao, L. et al. Repeated Growth and Bubbling Transfer of Graphene with Millimetre-Size Single-Crystal Grains using Platinum. *Nat. Commun.* **3**, 699 (2012).

16. Kim, G. et al. Growth of High-Crystalline, Single-Layer Hexagonal Boron Nitride on Recyclable Platinum Foil. *Nano Lett.* **13**, 1834–1839 (2013).

**Acknowledgments:**

T.A.Chen, C.P.C., H.S.W. and L.J.L. thank the support from Taiwan Semiconductor Manufacturing Company (TSMC). W.H.C. acknowledges the supports from the Ministry of Science and Technology of Taiwan (MOST-108-2119-M-009-011-MY3, MOST-107-2112-M-009-024-MY3) and from the CEFMS of NCTU supported by the Ministry of Education of Taiwan. Y. Z. gratefully acknowledges the financial support from the National Natural Science Foundation of China (Nos. 51861135201). Q. F. thanks the National Natural Science Foundation of China (No. 21688102, and No. 21621063), and Strategic Priority Research Program of the Chinese Academy of Sciences (Grant No. XDB17020000) for financial supports. B.I.Y. acknowledges support by the US Department of Energy (DE-SC0012547) and a stimulating discussion with Tony Ivanov (US Army Research Laboratory). T.A.Chen and L.J.L. acknowledge useful discussions with Dr. Steven Brems at Imec.

**Author statement:**

**Methods:**

**Growth of Cu (111) thin films.** The as-received sapphire substrates were first etched in a mixed $H_2SO_4$: $H_3PO_4$ aqueous solution at 300 °C for 20 min. The etched sapphire substrates were cleaned by immersing into ultra-pure water for 5 min. After cleaning, the sapphire substrates were loaded into a sputtering chamber for Cu deposition. The chamber was maintained at room temperature under Ar pressure of 0.3 mTorr, giving rise to a Cu deposition rate of 2 nm/sec. (10)

**Chemical Vapor Deposition of *h*BN.** The monolayer *h*BN films were grown in a 3-inch furnace tube with three heating zones using low-pressure chemical vapor deposition (LPCVD). The 2-inch Cu (111)/sapphire substrate was placed at the center heating zone of the main chamber. Ammonia borane (97%, ~60 mg) was used as the precursor and loaded into a sub-chamber at the upstream side of the main chamber. The furnace was first pumped down to a base pressure of 5.0 Torr. Before growth, the substrate was annealed at 1,050 °C for 60 min under a $H_2$ gas flow of 300 sccm. Then the sub-chamber (with precursors) was heated to 85 °C by using heating belt and kept for 30 min, followed by introducing the precursor into the main chamber for 30 min to grow *h*BN on the Cu (111)/sapphire substrates. After *h*BN growth, the sub-chamber was closed, and the main chamber was naturally cooled down to room temperature under a $H_2$ gas flow of 30 sccm.

**Transfer of *h*BN films onto arbitrary substrates.** The as-grown monolayer *h*BN film was detached from the Cu (111)/sapphire substrate by electrochemical delamination. A poly (methyl methacrylate) (PMMA) film was first spin-coated on the as-grown *h*BN/Cu (111)/sapphire as a protection layer. Then a thermal release tape (TRT; #3195M) was covered on the PMMA/*h*BN/Cu (111)/sapphire to avoid possible folding during the transfer process. The electrochemical delamination was performed in an aqueous solution of NaOH (1M) as the electrolyte, using the Cu layer in the TRT/PMMA/*h*BN/Cu (111)/sapphire stack as the cathode and a Platinum foil as the anode with an applied dc voltage of 4 V. During this process, the TRT/PMMA/*h*BN stacked film was detached from the Cu (111)/sapphire with the help of the hydrogen bubble generation at the *h*BN/Cu interface. After the detachment, TRT/PMMA/*h*BN stacked film can be placed on the target substrate. The TRT can be released by baking the TRT/PMMA/*h*BN/substrate on a hot-plate at 180 °C. The PMMA film was finally removed by immersing the sample in hot acetone for 40 min, leaving behind a monolayer $h$BN film on the target substrate.

**Chemical Vapor Deposition of MoS$_2$.** The highly oriented monolayer MoS$_2$ were grown on sapphire substrates by chemical vapor deposition in a horizontal hot-wall 3″ furnace tube with two heating zones. The maximum substrate size can be up to 2″ wafer. High-purity S (99.5%, Alfa) and MoO$_3$ (99%, Aldrich) powders were used as the reaction precursors. The S powder was placed in the front heating zone at the upstream side of the furnace, and the temperature was maintained at 140 °C during the reaction. The MoO$_3$ powder was put into a quartz boat in the center heating zone of the furnace. The temperature of the center heating zone was gradually ramped to 740 °C and held for 5 min. During this process, MoS$_2$ were grown on the sapphire substrates placed at the downstream side of the MoO$_3$ quartz boat. All growths were performed in Ar flowing gas (90 sccm) at a base pressure of 30 Torr. Finally, the furnace was naturally cooled down to room temperature. (17)

**Transfer of Monolayer MoS$_2$.** After CVD growth, the monolayer MoS$_2$ films on sapphire were transferred onto a target substrate using PMMA. A PMMA was first spin-coated on the as-grown MoS$_2$/sapphire as a protection layer. Then a thermal release tape (TRT; #3195M) was covered on the PMMA/MoS$_2$/sapphire to avoid possible folding during the transfer process. The PMMA/MoS$_2$/sapphire stacked film were then immersed in a NH$_4$OH solution (NH$_4$OH: DI water = 17:100) at 100 °C for 20 min. Then, the detached TRT/PMMA/MoS$_2$ stacked films were immersed into deionized water to dilute etchant and residues. Then, TRT/PMMA/MoS$_2$ stacked film can be placed on the target substrate, the TRT can be released by baking on a hot-plate at 180 °C and the PMMA film was finally removed by immersing the sample in hot acetone for 40 min, leaving behind a monolayer MoS$_2$ film on the target substrate. (17)

**LEEM/PEEM/$\mu$-LEED characterizations.** The structural analysis of $h$BN was performed using low energy electron microscopy (LEEM) with a field emission gun, photoemission electron microscopy (PEEM) excited by a mercury lamp, and micro-spot low energy electron diffraction ($\mu$-LEED, with an area of about 3 $\mu$m in diameter). The system (Elmitec LEEM-III) is composed of a preparation chamber, a main chamber for imaging, and a deep ultraviolet laser. (18) The sample was first annealed in the preparation chamber in ultrahigh vacuum (1×10$^{-9}$ Torr) at 600 °C for 6 hr and then transferred to the main chamber for LEEM/PEEM/$\mu$-LEED experiments.

**STM characterizations.** An Omicron ultrahigh vacuum (UHV) VT-STM system was utilized for the atomic-scale structural characterization with a base pressure better than 10$^{-10}$ mbar. The $h$BN/Cu sample was annealed at 900 K for 5.5 hr before the STM measurements. All the STM images were captured under a constant current mode at room temperature. The atomic-scale morphology of $h$BN/Cu sample was directly analyzed by high-resolution STM.

**EBSD characterization.** EBSD was performed on JEOL JSM-7800F PrimeSEM with collection accessory from EDAX at a 20 KV accelerating voltage with the sample stage tilting at 70°.

**TEM characterization.** TEM image was obtained from transferred $h$BN/SiO$_2$/Si with an acceleration voltage of 200 keV in FEI Tecnai Osiris Transmission Electron Microscope.

**AFM characterization.** AFM data were acquired in a tapping mode over scan area using a silicon tip.

**XPS Measurement.** The X-ray photoemission spectroscopy (XPS) spectra were performed in a Perkin Elmer PHI 5400 system which is equipped with hemispherical analyzer with an overall resolution of 0.05 eV. Energy span and energy scale is calibrated by setting the 4f 7/2 line of gold and the 2p 3/2 line of copper to 84 and 932.67 eV, respectively. The XPS spectra were measured from the as-grown monolayer $h$BN film on Cu (111)/sapphire. The B-1s and N-1s emission peaks confirm the formation of $h$BN. The binding energy of B1s and N1s is located at 190.4eV and 398eV, respectively. The B:N atomic ratio is calculated from the integrated intensities of these peaks, which yields a B:N ratio of 1:1.03, indicative of a good stoichiometry for the CVD-grown monolayer $h$BN films.

**Raman Measurement.** The Raman measurements were performed on transferred $h$BN film using a 532-nm solid state laser as the excitation source. The excitation light with a power of 2.5 mW was focused onto the sample by a ×100 objective lens (N.A. = 0.9). The signal was collected by the same objective lens, analyzed by a 0.75-m monochromator and detected by a liquid-nitrogen-cooled CCD camera. The $E_{2g}$ band of $h$BN is located at 1371 cm$^{-1}$ and its full width at half maximum (FWHM) value is 14.5 cm$^{-1}$, confirming that the transferred $h$BN film is monolayer thick with high crystalline quality.

**XRD Measurement.** X-ray diffraction (XRD, Bruker D8-Discover) $\theta$-$2\theta$ and 360° azimuthal ($\varphi$) scans were conducted using a Cu K$\alpha$ radiation source ($\lambda$=1.54 Å). XRD azimuthal $\varphi$ scan of the twined Cu (111) films was operated at $\chi$= 70.5°, $\omega$= 16.2° and $2\theta$= 43.3°, XRD azimuthal $\varphi$ scan of the untwined Cu (111) films was operated at $\chi$=70.5°, $\omega$= 21.5° and $2\theta$= 43.3°.

**First-principles Calculation.** The first-principles calculations were carried out based on density function theory (DFT) as implemented in the Vienna Ab-initio Simulation Package (VASP)(19) within the MedeA.(20) The projector augmented wave method, and the exchange-correlation potential described by the PBE-GGA,(21) and the vdW correction vdW-DF (optB86b) functionals(22) is used to calculate the distance and binding energy between BN molecule and copper substrates. For plane-to-plane and edge-to-step epitaxy model, a rigid $B_6N_7$ molecule is physically absorbed on a 6×6 and 6×7$\sqrt{3}$ Cu (111) surface with four copper layers thickness with 2 ×2 × 1 and 2 ×1 × 1 k-grid, respectively. To eliminate the spurious interaction due to slab model of supercell, the vacuum thickness is larger than 17 Å, the in-plane distance of $B_6N_7$ molecule away from another side of step edge is at least larger than 10 Å, and the energy cutoff of plane waves is 400 eV. The lattice constant of Cu (111) used in simulation is a=2.6 Å at the reaction temperature 1,000 °C, with $h$BN assuming the same lattice constant for simplification. The interlayer distance of Cu (111) is 2.05 Å. The height of BN molecule from copper surface is optimized until the change of the energy and the force reach $10^{-4}$ eV and $10^{-2}$ eV/ Å, respectively. For plane to plane epitaxy, the optimized vertical distance from top layer of copper surface, d, of six configurations are d=1.98 Å for $N_IB_{II}$, d=1.77 Å for $N_{II}B_{III}$, d=2.03 Å for $N_{III}B_I$, d=1.98 Å for $N_IB_{III}$, d=1.77 Å for $N_{III}B_{II}$, and d=2.03 Å for $N_{II}B_I$. The calculated binding energies, defined by -$E_b$= $E_{B_6N_7-Cu}$ - $E_{B_6N_7}$ - $E_{Cu}$ where $E_{B_6N_7-Cu}$ is the total energy of $h$BN flake-Cu (111) system, $E_{B_6N_7}$ and $E_{Cu}$ are the energies of the $B_6N_7$ flake and Cu substrate (with or without step edge), of 6 total stacking structures in the presence of a step are plotted by varying the in-plane distance $D_i$ to the step-edge. The six lowest energetic configurations in the presence of a step are located at $D_i$ = 2a/$\sqrt{3}$ for $N_IB_{II}$ (d=1.99 Å), 4a/$\sqrt{3}$ for $N_{II}B_{III}$ (d=1.77 Å), 1.5a/$\sqrt{3}$ for $N_{III}B_I$ (d=2.08 Å), 2.5a/$\sqrt{3}$ for $N_IB_{III}$ (d=1.98 Å), 1.5a/$\sqrt{3}$ for $N_{III}B_{II}$ (d=1.84 Å), and 2a/$\sqrt{3}$ for $N_{II}B_I$ (d=2.06 Å), as shown in **Extended Data Fig. 8**. Note that when $D_i$ is less than $\sqrt{3}\,a/3$ (too close to the step-edge), the energy dramatically increases.

We study the energy difference between $N_IB_{III}$ (0°) and $N_IB_{II}$ (60°), two lowest energy structures docking to A-step and B-step edges, with different length of docking contact. We construct a model of BN *stripe* composed of aromatic rings docking to the step edge with length. We find that the energy difference of the two configurations $N_IB_{III}$ (0°) and $N_IB_{II}$ (60°), an indicator of selectivity for mono-orientation, increases rapidly exponentially with the docking contact, and approaches to ~0.78 eV (corresponding at $k_BT$ ~0.11eV at 1300 K), amplifying the Boltzmann selectivity factor for just 5-6 hexagons contact to above $10^3$ (see **Extended Data Fig. 9a-b**). The large energy difference means that the step-edge plays a role of "orientation filter" that allows a predominant phase of BN during the growth. Such an energy difference apparently ensures the mono-orientation growth. Factually, the competing subcritical nuclei must be larger and, importantly, likely elongated with the edge for more energetically favorable contact. Obviously such energy value roughly scales with the contact length C and the selectivity factor very rapidly increases with size, as $e^C$.

The effect of misfit and misalignment of BN molecule to the step-edge is studied by calculating the binding energy of a $B_7N_7$ molecule at 3-hexagons contact length with small tilt-angle along the step-edge (the aligned structure with zero tilt-angle corresponds to $N_IB_{III}$ and $N_IB_{II}$). The lattice constant of Cu (111) and *h*BN are allowed to be different as $a_{Cu111}$=2.6 Å and $a_{hBN}$=2.5 Å, with a 3.8 % lattice misfit. We find that although the total binding energy is weakened due to lattice misfit (the weaker interaction of plane-to-plane epitaxy), the energy difference between 0° and 60° orientations remains nearly the same (from 0.3 eV to 0.27 eV). The calculated binding weakens as the tilted angle becomes larger, indicating the most stable configuration at the initial nucleation growth when docking is tight, well aligned to the edge-step, see **Extended Data Fig. 9c-d**. The result suggests that single orientated *h*BN was epitaxially grown on Cu (111) surface which was guided by mostly steps edge, and the moiré was formed to release the strain between Cu (111) and *h*BN (lattice mismatch or surface topography) after large area film are covered on surfaces. Particularly, the cooling after *h*BN large-area growth shall result in strains which may be released by local straining of *h*BN and moiré pattern formation. Therefore, we believe that moiré is not actually affecting the growing process, but occurs afterwards.

**Method References** (*17-25*)

17. Hsu, W.-F. et al. Monolayer MoS$_2$ Enabled Single-Crystalline Growth of AlN on Si(100) Using Low-Temperature Helicon Sputtering. *ACS Appl. Nano Mater.* **2**, 1964-1969 (2019).

18. Jin, L., Fu, Q., Mu, R., Tan, D. & Bao, X. Pb intercalation underneath a graphene layer on Ru(0001) and its effect on graphene oxidation. *Phys. Chem. Chem. Phys.* **13**, 16655–16660 (2011).

19. Kresse, G. & Furthmüller, J. Efficiency of ab-initio total energy calculations for metals and semiconductors using a plane-wave basis set. *Comput. Mater. Sci.* **6**, 15–50 (1996).

20. Materials Exploration and Design Analysis. Materials Design, Inc; Angel Fire, NM, USA: (2016).

21. Perdew, J. P., Burke, K. & Ernzerhof, M. Generalized gradient approximation made simple. *Phys. Rev. Lett.* **77**, 3865–3868 (1996).

22. Klimeš, J., Bowler, D. R. & Michaelides, A. Van der Waals density functionals applied to solids. *Phys. Rev. B* **83**, 195131 (2011).

23. Yankowitz, M. et al. Emergence of superlattice Dirac points in graphene on hexagonal boron nitride. *Nature Physics* **8**, 382-386 (2012).

24. Joshi, S. et al. Boron nitride on Cu (111): an electronically corrugated monolayer. *Nano Letters* **12**, 5821-5828 (2012).

**Extended data figure legends**

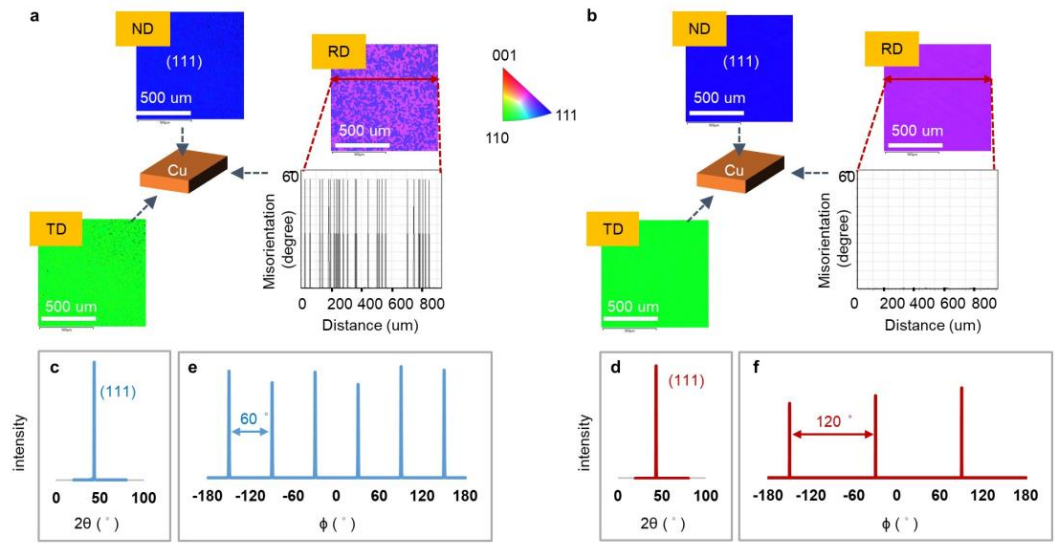

**Extended Data Fig. 1 | Analysis of Cu (111) crystal orientation on c-sapphire substrates. a**, EBSD inverse pole figure (IPF) mapping (1mm x 1mm) of the annealed Cu substrate at 1,000 °C for 1 hr, the normal direction (ND) and rolling direction (RD) mappings indicate Cu (111) is polycrystal with in-plane rotation by 60° (misorientation of line scan). **b**, IPF map of the Cu substrate annealed at 1,050 °C for 1 hr, where the film is characterized as single-crystal Cu (111), and no twined grain is founded. **c**, XRD $\theta$-$2\theta$ scan of the annealed Cu (111)/c-sapphire substrate at 1,000 °C for 1 hr and **d**, annealed at 1,050 °C for 1 hr, a Cu (111) peak at $2\theta = 43.3°$. **e**, XRD $\varphi$ scan of the annealed Cu (111)/c-sapphire substrate at 1,000 °C for 1hr with an in-plane 60° rotation and **f**, annealed at 1,050 °C for 1 hr with an in-plane 120° rotation. Note that the 1,050 °C annealed sample shows the signature of single-crystal without in-plane misorientation because *h*BN triangle is with $C_3$-symmetry and it is symmetric after 120° rotation.

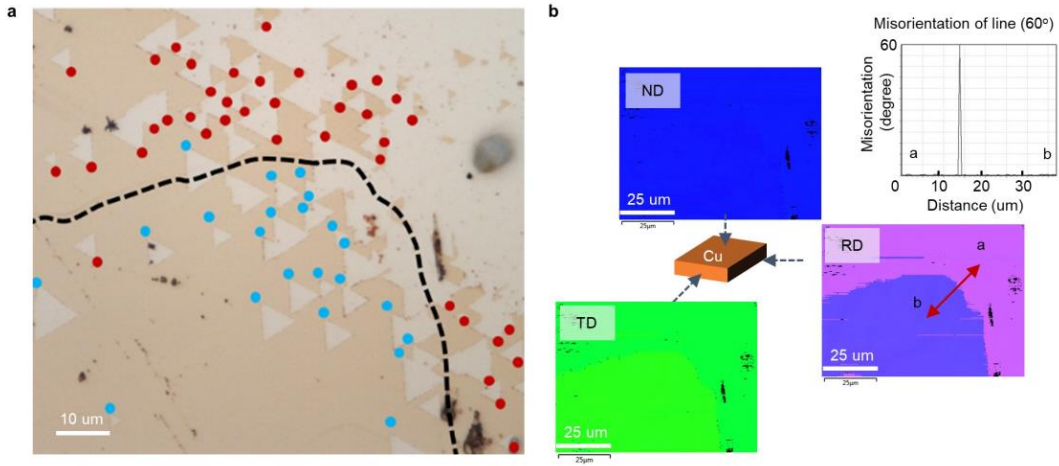

**Extended Data Fig. 2 | Single-oriented *h*BN flakes in different Cu grains. a**, Optical microscope image of the *h*BN grown on different Cu (111) grains (black dashed lines indicate the grain boundary at the adjacent twin grains). Oppositely-oriented *h*BN flakes were marked by red and blue circles. **b**, EBSD IPF maps of the area shown in **a**. The misorientation of line scan indicates the twin grain is rotated by 60° (grain a → grain b). The RD map clearly shows the in-plane orientation difference in grain a and b.

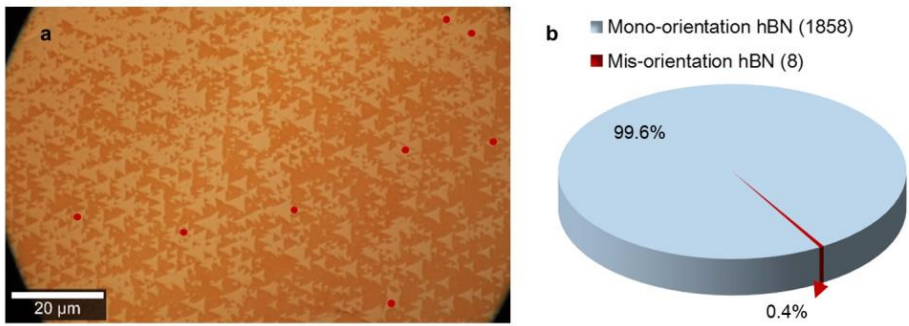

**Extended Data Fig. 3 | Statistical analysis for the orientation distribution of *h*BN triangular flakes. a**, Optical microscope image of *h*BN grown on Cu (111)/c-sapphire substrate at 1,050 ºC. Misaligned *h*BN flakes are marked by red circles. **b**, A statistical analysis on the OM image shown in a, where > 99.6% *h*BN flakes are aligned to one direction on Cu (111).

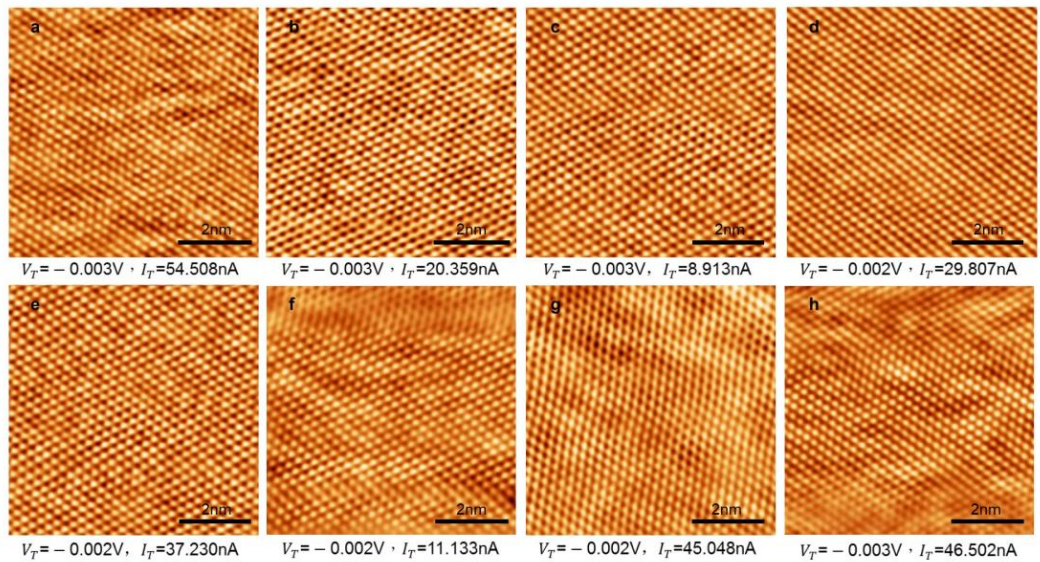

**Extended Data Fig. 4 | STM images from randomly selected locations. a-h,** the images are taken in a 1.5 × 1.5 cm² hBN film on Cu (111)/c-sapphire, where the angle versus horizontal line is 26.5° ± 1°, indicating that the hBN is a single-crystal film.

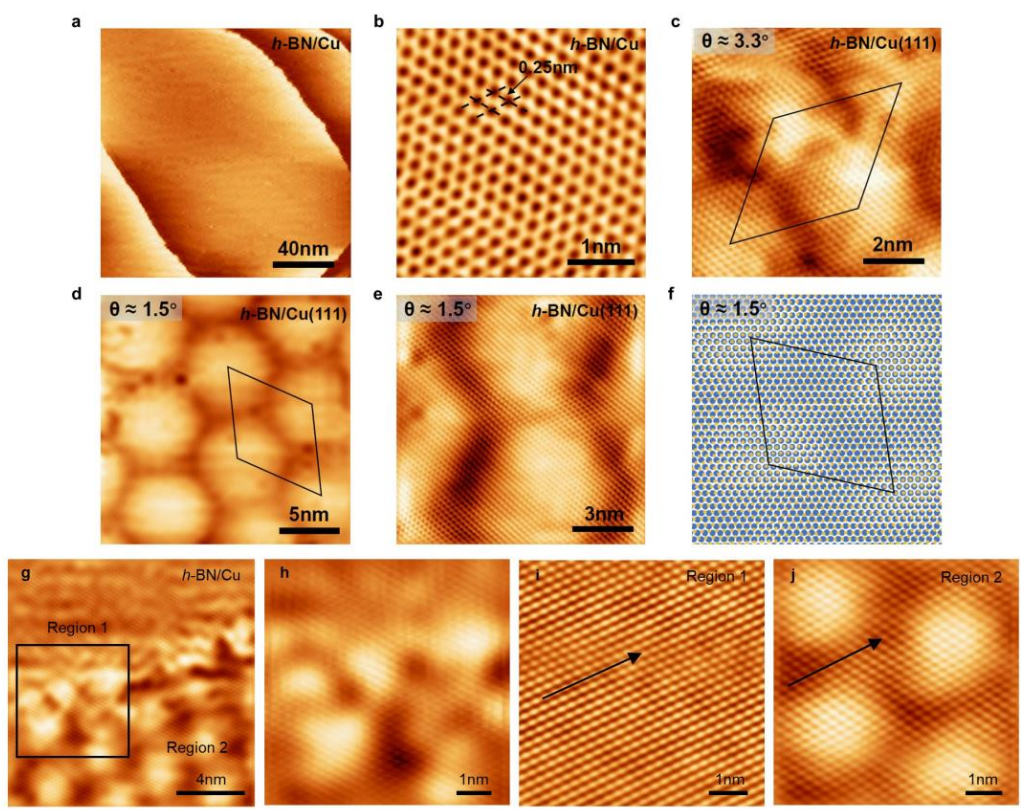

**Extended Data Fig. 5 | STM characterizations of the atomic structure of monolayer hBN on Cu (111) substrate. a,** Large-scale STM image ($V_T$ = -1.008 V, $I_T$ = 3.90 nA, T = 300 K) of hBN/Cu. **b,** Atomic-scale STM image ($V_{Tip}$ = -0.003 V, $I_{Tip}$ = 46.50 nA, T = 300 K) of hBN/Cu. **c,**

Typical STM image ($V_{Tip}$ = -0.003 V, $I_{Tip}$ = 8.91 nA, T = 300 K) of hBN on Cu (111) with a relative rotation angle of θ ~3.3°, showing a moiré pattern with a period of 4.20nm. **d**, Typical STM image ($V_{Tip}$ = -0.032 V, $I_{Tip}$ = 18.51 nA, T = 300 K) of hBN on Cu (111) with a relative rotation angle of θ~ 1.5°, showing a moiré pattern with a period of 7.75nm. The unit cell of the moiré pattern is highlighted by a black rhombus. **e**, Magnified STM image ($V_{Tip}$ = -0.039 V, $I_{Tip}$ = 18.51 nA, T = 300 K) of hBN on Cu (111) with a relative rotation angle of θ ~ 1.5°. **f**, Simulation of the moiré pattern for monolayer hBN on Cu (111) with a relative rotation angle of θ ~ 1.5°. The unit cell of the moiré pattern for hBN/Cu (111) is highlighted by a black rhombus. The large-scale STM image in Extended Data Fig. 5a shows a large-area flat terrace of hBN/Cu with a clean surface. Further atomic-scale STM image of hBN/Cu (Extended Data Fig. 5b) presents a honeycomb structure with a lattice constant ~0.25 nm, which coincides well with the lattice parameter of hBN. Notably, in some typical regions of hBN/Cu (111), several kinds of moiré patterns with different periods were observed. For instance, Extended Data Fig. 5d and 5e show a moiré pattern with a period of ~7.75 nm, and Extended Data Fig. 5c shows another moiré pattern with a period of ~4.20 nm. Such moiré patterns arise from the lattice mismatch and/or relative rotation between hBN and underlying Cu (111) substrate, and the moiré periods (D) are correlated with the relative rotation angles (θ) between hBN and Cu (111) lattice by the following equation (23):

$$D = \frac{(1+\delta)a}{\sqrt{2(1+\delta)(1-\cos\theta) + \delta^2}}$$

Where δ is the lattice mismatch (~2%) between hBN and Cu (111) lattice (24), and *a* is the lattice constant of hBN. Consequently, the θ for the moiré pattern with D ~7.75 nm (Extended Data Figs 5d and 5e) was calculated as ~ 1.5°, and the simulated moiré pattern generated from monolayer hBN stacking on Cu (111) with a rotation angle of θ ~ 1.5° (Extended Data Fig. 5f) fits well with the STM result (Extended Data Figs. 5d and 5e). Besides, the θ for the moiré pattern with D ~4.20 nm was calculated as ~ 3.3° (Extended Data Fig. 5c).

Extended Data Fig. 5 **g-j**. shown STM images for the boundary between the areas with and without moiré pattern. **g**, Typical STM image ($V_{Tip}$ = -0.003 V, $I_{Tip}$ = 8.10 nA, T = 300 K) of hBN/Cu at the boundary site. **h**, Magnified STM image ($V_{Tip}$ = -0.003 V, $I_{Tip}$ = 8.10 nA, T = 300 K) of the boundary from figure a (highlighted by the black square) showing that the hBN lattices present a perfect coherence at the boundary site. **i**, Magnified STM image ($V_{Tip}$ = -0.003 V, $I_{Tip}$ = 17.93 nA, T = 300 K) of hBN/Cu without moiré pattern in region 1 of a. **j**, Magnified STM image ($V_{Tip}$ = -0.003 V, $I_{Tip}$ = 10.78 nA, T = 300 K) of hBN/Cu with moiré pattern in region 2 of a. The hBN atomic rows in the areas are along the same direction, as indicated by the black arrows.

The STM image in Extended Data Fig. 5g was captured at a typical boundary region featured with moiré (region 2) and non-moiré (region1) areas. Magnified atomic resolution image at the boundary site in Extended Data Fig. 5h demonstrates that, the hBN presents perfect lattice coherence at the patching boundary. The hBN atomic rows in the two adjacent area are along the same direction, as shown in Extended Data Fig. 5i-j. All these images indicate that, the hBN is aligned well and has mono-orientation. These results indicate an epitaxial growth behavior of hBN on Cu (111), and the formation of moiré patterns does not affect the hBN orientation. We believe that the hBN complete the single-crystal growth at high temperatures and the strain associated with the sample cooling after growth results in the formation of local moiré pattern.

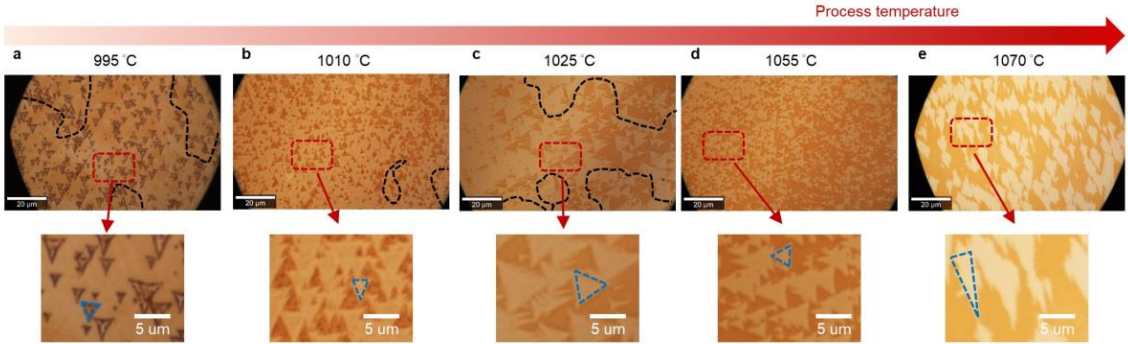

**Extended Data Fig. 6 | Optical microscope image of the *h*BN grown on Cu (111)/c-sapphire substrate at different temperatures.** All as-growth *h*BN/Cu substrate was slowly oxidized at 150 °C in air to enhance contrast between *h*BN and Cu film and the black dashed lines indicate Cu twin grain boundaries. **a**, *h*BN grown at 995 °C and the triangles were aligned in the same Cu grain. Note the flakes are easily oxidized due to their poor quality. **b**, *h*BN grown at 1,010 °C. Note that the *h*BN flakes grown at 995 and 1,010 °C are easily damaged after 150 °C oxidation in air. **c-d**, All *h*BN triangles pointed to the same direction in the same Cu (111) grain at 1,025 ~1,055 °C and triangles seem unchanged after oxidation. **e**, *h*BN flakes all aligned in the same direction with different domain shapes (stretched) at high temperature (1,070 °C).

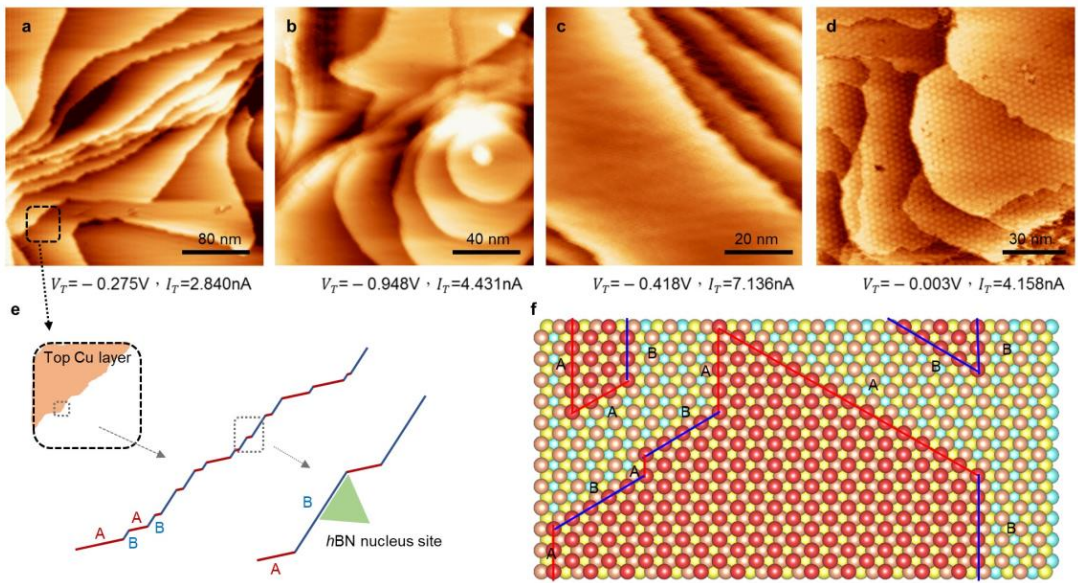

**Extended Data Fig. 7 | STM images of *h*BN/Cu (111). a-c**, STM images of *h*BN/Cu steps without moiré pattern and **d**, with moiré pattern. These images indicate that step edges are often observed in our Cu (111) crystals. **e,** A schematic diagram illustrates that the meandering steps consist of segments of A and B types, and BN is kinetically nucleated at the A to B corner while docking to stronger binding sites, B-steps, with proper orientation. **f**, An atomic model of step edges on top Cu (111) surface, two kind of step edge termination A-, and B-step are identified.

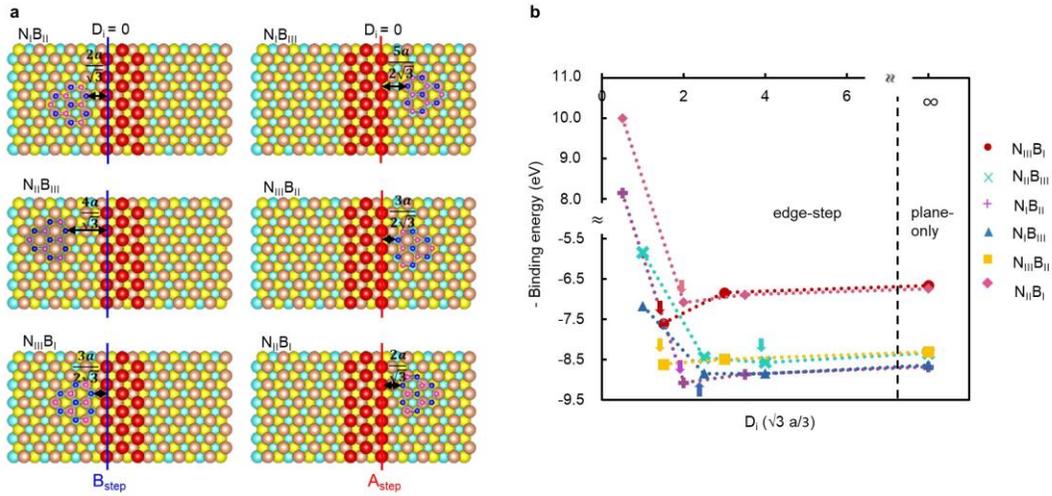

**Extended Data Fig. 8 | Calculated binding energy of six typical B$_6$N$_7$-Cu (111) configurations with the consideration of top-layer Cu step edge docking. a,** The atomic model of six configurations with optimized in-plane distance D$_i$ away from A-step (red) or B-step (blue) edge. **b,** The binding energy with the same stacking varying as a function of D$_i$ (in unit of $\sqrt{3}$ a/3, where a is the lattice constant of Cu (111)), where the lowest energy configuration is labeled by arrow. The plane-to-plane epitaxy corresponds to the absence of (or infinity distance away from) the step-edge.

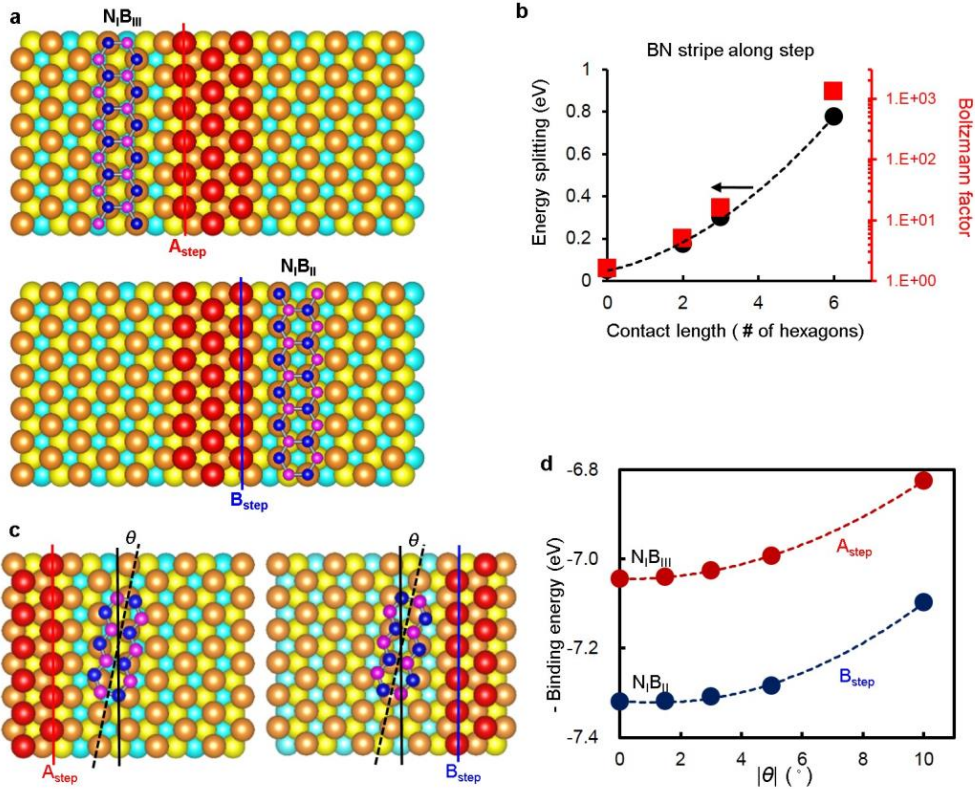

**Extended Data Fig. 9 | Comparison of planar epitaxy and step-edge docking of *h*BN on Cu (111) surfaces. a**, Atomic model of fully docking configurations of $N_IB_{II}$ (0°) and $N_IB_{III}$ (60°). **b**, Calculated binding energy difference (left axis, black circle) and corresponding Boltzmann factor (right axis, red square) between $N_IB_{II}$ and $N_IB_{III}$ of BN *stripe*-Cu (111) configurations as a function of docking length to the top-layer Cu step edge. **c-d,** Calculated energy of misfit $B_7N_7$-Cu (111) configurations as a function of small tilted angle along with top-layer Cu step edge docking. **c,** Atomic models of $B_7N_7$-Cu (111) configurations docking to A-step or B-step at tilt angle $\theta$. The aligned structures ($\theta = 0°$) correspond to $N_IB_{II}$ and $N_IB_{III}$ (two lowest energetic docking structures in the vicinity of A-step and B-step edges). The lattice misfit between *h*BN and Cu (111) is 3.8 % ($a_{hBN}$=2.5 Å, and $a_{Cu (111)}$=2.6 Å) **d,** The binding varies as small tilted angle along the edge steps.

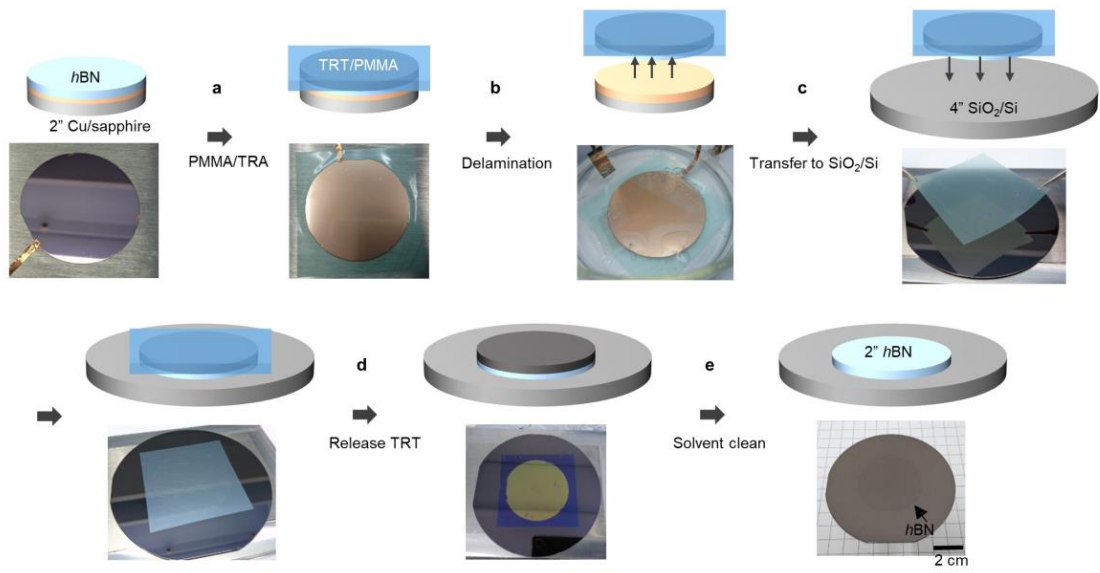

**Extended Data Fig. 10 | Schematics and photos illustrating the *h*BN transfer processes. a**, A poly(methyl methacrylate) (PMMA) film was first spin-coated on the as-grown *h*BN/Cu (111)/sapphire as a protection layer. Then a thermal release tape (TRT; #3195M) was covered on the PMMA/*h*BN/Cu (111)/sapphire to avoid folding during the transfer process. **b**, The electrochemical delamination was performed in an aqueous solution of NaOH (1M) as the electrolyte, the Cu layer in the TRT/PMMA/*h*BN/Cu (111)/sapphire stack as the cathode, and a Platinum foil as the anode with an applied dc voltage of 4V. During this process, the TRT/PMMA/*h*BN stacked film was detached from the Cu (111)/sapphire through the generation of hydrogen bubbles at the *h*BN/Cu interfaces. **c**, TRT/PMMA/*h*BN stacked film placed on the 4 inch SiO$_2$/Si substrate. **d**, TRT can be released by baking the TRT/PMMA/*h*BN/substrate on a hot-plate at 180 °C. **e**, The PMMA film was removed by immersing the sample in hot acetone for 40 min, leaving behind a 2" monolayer *h*BN film on the 4 inch SiO$_2$/Si wafer.

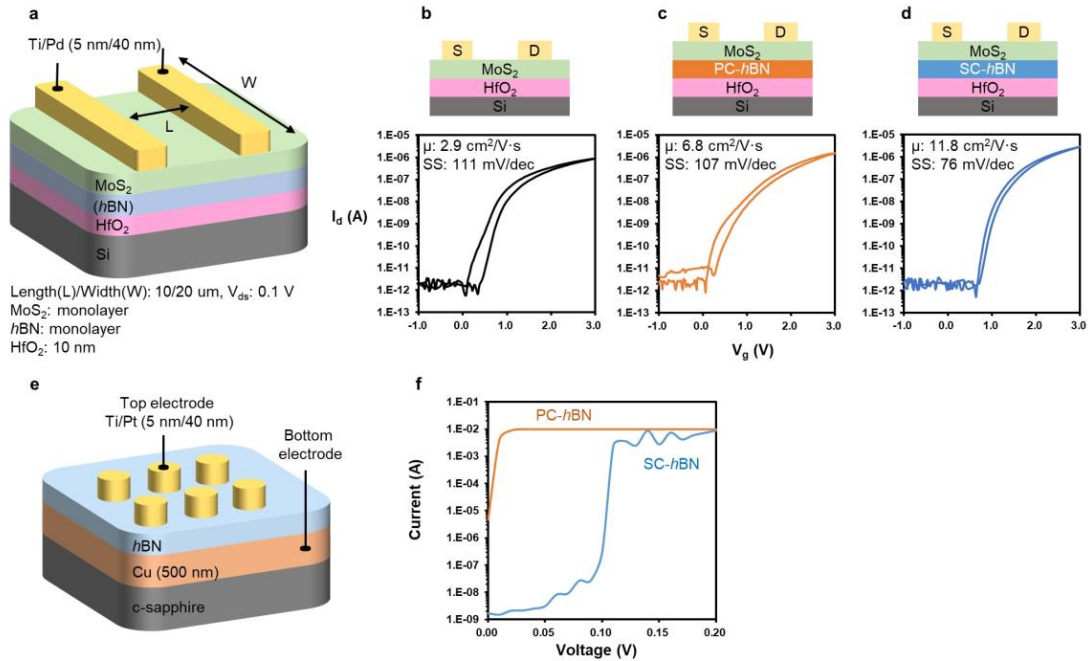

**Extended Data Fig. 11 | FETs built with and without *h*BN as an interface dielectric in a bottom-gate configuration. a-d**, Schematics of various $MoS_2$ FET device structures examined and transfer characteristics ($I_d$–$V_g$) of $MoS_2$ FETs at $V_{ds}$ = 0.1 V based on the device structures.

The Extended Data Fig. 11 shows the transfer characteristics (drain current $I_d$ – gate voltage $V_g$) of monolayer $MoS_2$ field-effect transistors (FETs) at a driving voltage of $V_{ds}$ = 0.1 V. Monolayer *h*BN and $MoS_2$ are synthesized by CVD method. The *h*BN film was transferred on a Si substrate with a 10-nm $HfO_2$ dielectric layer, followed by a single layer $MoS_2$ on top. The metal contacts were processed by photolithography, and evaporated with Ti (5 nm) and Pd (50 nm) by an e-gun evaporator. The schematics of a typical monolayer $MoS_2$ FET on $HfO_2$ (Extended Data Fig. 11b), a $MoS_2$ monolayer on a polycrystalline (PC-) *h*BN monolayer (Extended Data Fig. 11c), a $MoS_2$ monolayer on a single-crystal (SC-) *h*BN monolayer (Extended Data Fig. 11d) are shown. The extracted two-probe electron mobility at room temperature in vacuum are 2.9, 6.8 and 11.8 $cm^2$/V·s, respectively. The mobility of $MoS_2$ is improved significantly (about an order of magnitude) after replacing the polycrystalline *h*BN by a single-crystal *h*BN. A suppressed current hysteresis has also been observed when using a single-crystal *h*BN as the buffer layer. The results indicate that single-crystal *h*BN can be used to enhance the electrical performance of 2D-based transistors.

Extended Data Fig. 11e shows a schematic of *h*BN metal-insulator-metal (MIM) structure used to examine the quality of *h*BN. **f**, I−V curves for the MIM tunnel junctions with either a polycrystalline (PC, orange) or a single-crystal (SC, blue) *h*BN monolayer sandwiched between Pt/Ti (top) and Cu (bottom) electrodes.

Electrical contacts were fabricated by photolithography and e-gun evaporation of Ti (5 nm) and Pt (40 nm) to form 100×100 μm² pads on as grown *h*BN/Cu/c-sapphire substrate. Extended Data Fig. 11f shows the characteristic I−V curves for Pt-Ti/BN/Cu devices with different *h*BN insulating layer. The device with single-crystal monolayer *h*BN exhibits a large breakdown voltage (~0.1 V), whereas the device with polycrystalline monolayer $h$BN shows a direct tunneling characteristics.